\documentclass[aps,pre,twocolumn,showpacs,floatfix,superscriptaddress, longbibliography]{revtex4-1}

\usepackage{amsmath,amssymb,graphicx,color,braket,hyphenat,makeidx,xcolor,dsfont,subfigure,blindtext}
\usepackage[ocgcolorlinks,colorlinks=true,linkcolor=blue,citecolor=red,linktocpage=true]{hyperref}

\newcommand{\blue}{\color{\blue}}

\newcommand{\tr}{\mathrm{Tr}}

\def\id {{\mathds 1}}

\begin{document}

\title{Synchronization along quantum trajectories}

\author{Najmeh Es'haqi-Sani} 
\affiliation{International Centre for Theoretical Physics ICTP, Strada Costiera 11, I-34151, Trieste, Italy}
\affiliation{Department of Physics, Ferdowsi University of Mashhad, Mashhad, PO Box 91775-1436, Iran}

\author{Gonzalo Manzano}
\affiliation{International Centre for Theoretical Physics ICTP, Strada Costiera 11, I-34151, Trieste, Italy}
\affiliation{Scuola Normale Superiore, Piazza dei Cavalieri 7, I-56126, Pisa, Italy}

\author{Roberta Zambrini}
\affiliation{Institute for Cross-Disciplinary Physics and Complex Systems IFISC (UIB-CSIC), Campus Universitat Illes Balears, E-07122 Palma de Mallorca, Spain}

\author{Rosario Fazio}
\affiliation{International Centre for Theoretical Physics ICTP, Strada Costiera 11, I-34151, Trieste, Italy}
\affiliation{Dipartimento di Fisica, Universit\`a di Napoli "Federico II", Monte S. Angelo, I-80126 Napoli, Italy}

\begin{abstract}
We employ a quantum trajectory approach to characterize synchronization and phase-locking between open quantum systems in nonequilibrium steady states. We exemplify our proposal for the paradigmatic case of two quantum Van der Pol oscillators interacting through dissipative coupling. We show the deep impact of synchronization on the statistics of phase-locking indicators and other correlation measures defined for single trajectories, spotting a link between the presence of synchronization and the emergence of large tails in the probability distribution for the entanglement along trajectories. Our results shed new light on fundamental issues regarding quantum synchronization providing new methods for its precise quantification.
\end{abstract}


\maketitle 

\section{Introduction}

Synchronization is one of the most universal manifestations of emergent 
cooperative behavior, observed in a broad range of physical, chemical and 
biological systems~\cite{Synch1,Synch2}. It can arise spontaneously as a progressive 
adjustment of rhythms between oscillatory units due to their weak interaction 
and despite their different natural frequencies. Appealing examples with 
interesting applications comprise synchronization between hearth cardiac 
pacemaker cells \cite{Synch2}, chaotic laser signals~\cite{Colet} or 
micro-mechanical oscillators~\cite{Micro1,Micro2,Micro3}.

In the last decade, the interest in this paradigmatic phenomenon has been 
extended to the quantum realm, see e.g. Refs.~\cite{ZhirovEPJD, Goychuk, ZhirovPRL, Giorgi, 
Manzano, Fazio, Ludwig, LeePRL, Nazarov, Zueco, Holland, Walter, Rey, Lorch, 
Cabot, Sonar, Tilley, Solano, Karpat} on driven and spontaneous synchronization. Quantum mechanics plays a crucial role 
when exploring this phenomenon beyond the classical regime ~\cite{Review} and in relation to
the degree of synchronization that systems can reach~\cite{Fazio}. 
Quantum synchronization can be characterized with different outcomes ~\cite{Cabot2} using local or global indicators
in the system observables ~\cite{Review}. 
It has been shown that the emergence of this phenomenon is often connected to the generation of 
quantum correlations such as discord~\cite{Giorgi, Manzano, Manzano2, Giorgi2, 
Bemani} or entanglement~\cite{ZhirovPRB, Manzano, LeePRE, MarcTimme, Roulet, 
Lin}. 
However, a universal relation between quantum correlations and synchronization 
is not expected in general, and thus whether quantum synchronization may be used 
for witnessing quantum correlations is still an open question. 
In addition, quantum synchronization may also find applications for probing 
spectral densities in natural or engineered environments~\cite{GiorgiP, 
Nokkala}. 

In classical systems, spontaneous synchronization is usually characterized through the 
trajectories in phase-space ~\cite{Synch2}. In contrast, measuring synchronization in 
open quantum systems becomes more challenging and different avenues have been 
explored. For instance, temporal correlations in local observables can be quantified 
by using the Pearson correlation coefficient~\cite{Giorgi} or global quantum correlations
can be addressed through the synchronization error~\cite{Fazio}. 
Quantitative measures of phase-locking based on the expectation values of 
different non-local correlators~\cite{Fazio, Holland, Rey, Armour} have been proposed, but they are 
often not indicative of the underlying processes~\cite{Weiss}. 
Phase distributions computed from the Wigner quasi-probability distribution~\cite{LeePRL} or using phase states~\cite{Armour} have been used to gain extra insights in this context.
Finally, information measures of correlations like the mutual information~\cite{Ameri} or 
Renyi-entropies~\cite{Bastidas} have also been employed. 
In all these approaches, synchronization is computed through the expectation 
values of different (local or global) observables on the system density 
operator, as given by the solution of some suitable master equation.

In this paper we aim to go beyond the average effects of noise, and 
characterize synchronization along individual quantum trajectories in Hilbert 
space.  
The quantum trajectory approach describes the stochastic evolution of the pure 
state of the system of interest when environmental monitoring is 
available~\cite{milburn, trajectories}.
This formalism allows for a deeper notion of synchronization in the quantum 
regime, and enables one to explore a hidden link between the emergence of 
synchronization and the generation of entanglement along single stochastic 
realizations of the process, which cannot be inferred from the density 
operators. 

The impressive development of experimental techniques in the last decade allowed the generation and recording of 
quantum trajectories in a number of platforms, including ultrahigh-Q Fabry-Perot cavities~\cite{Haroche, Haroche2}, 
superconducting qubits~\cite{Murch, Roch, Devoret, Weber, Huard} and optomechanical systems~\cite{Aspelmeyer, Rossi}. 
Quantum trajectories have been used to detect phase transitions in the steady-state dynamics of dissipative quantum systems~\cite{GarrahanPRL}. Recently, Ref.~\cite{Weiss} provided a first clue on the potential of quantum trajectories in synchronization by using them to detect the presence of different phase-locking regimes. Here we aim to exploit at maximum the extra information that environmental measurements may offer us to give a deeper characterization of synchronization and phase-locking in the quantum regime.

We consider one of the most paradigmatic setups for the study of quantum synchronization, namely, a couple of (self-sustained) Van der Pol (VdP) oscillators weakly interacting through a dissipative coupling~\cite{LeePRE,Walter2}. 
The two VdP oscillators reach limit-cycles in the long time run, where phase locking may appear depending on the trade-off between the 
oscillators detuning and their coupling strength. We use the statistics of phase-locked trajectories as well as other natural indicators 
to study synchronization, therefore extending the concept to the single trajectory case. Synchronization may strongly manifest in the 
shape of the distribution of phase-differences and other synchronization indicators, whose variances drop in its presence. Even if our 
findings are mainly illustrated using a simple system of two quantum Van der Pol  oscillators, we expect our method to provide similar results in other setups.

\section{Model and quantum trajectories} \label{sec:model}

The VdP oscillator is a nonlinear dynamical system with two different dissipative contributions: a nonlinear damping term and a pumping term powering self-oscillations. This model has been largely studied in the context of synchronization and Hopf bifurcations of classical systems~\cite{Synch1, Synch2}. In the quantum case, the model of two quantum VdP oscillators interacting through dissipative coupling can be described with the help of the following Lindblad master equation ($\hbar=1$)~\cite{LeePRE,Walter2} 
\begin{align} \label{eq:master}
\dot{\rho} =& ~\mathcal{L}(\rho) = -i[H,\rho] + V \mathcal{D}[a_1 - e^{i \theta} a_2]\rho \nonumber \\ 
&+ \sum_{i=1}^2 \gamma_\downarrow^{(i)} \mathcal{D}[a_i^2]\rho + \gamma_\uparrow^{(i)} \mathcal{D}[a_i^\dagger]\rho,
\end{align}
where $\rho$ is the density operator of the two oscillators, $H=\sum_{i=1}^2 \omega_i a_i^\dagger a_i$ is the system Hamiltonian with frequency detuning $\Delta \omega \equiv \omega_2 - \omega_1$, and we denoted the dissipators as $\mathcal{D}[L]\rho = L \rho L^\dagger - \frac{1}{2}\{ L^\dagger L, \rho \}$ for any Lindblad operator $L$.
The positive rates $V$, $\gamma_\downarrow^{(i)}$ and $\gamma_\uparrow^{(i)}$, stand 
respectively for the coupling strength between oscillators, and the rates at which nonlinear damping and pumping processes occur. 
The angle $\theta$ will determine the phase difference between oscillators at which synchronization occurs.

The classical equations of motion for the oscillators amplitude are recovered for the annihilation operator expectations $\alpha_i = \langle a_i \rangle_{\rho} = \tr[a_i \rho]$ (first order moments) in the infinite photon limit $\gamma_\downarrow^{(i)} / \gamma_\uparrow^{(i)} \rightarrow 0$. 
The region of parameters $(\Delta \omega, V)$ for which phase-locking emerges for two VdP oscillators in the classical limit displays the usual
Arnold tongue V-shape centered around $\Delta \omega = 0$~\cite{LeePRE}. For symmetric local damping rates  $\gamma_{\uparrow, \downarrow}^{(1)} = \gamma_{\uparrow, \downarrow}^{(2)}$, 
it is simply given by $V = 2 |\Delta \omega|$~\cite{Walter2}.

On the contrary, the quantum limit is achieved when $\gamma_\downarrow^{(i)} / \gamma_\uparrow^{(i)} \rightarrow \infty$~\cite{LeePRL,LeePRE}. 
In this case the steady state solution $\pi$ of Eq.\eqref{eq:master}, obtained by solving $\mathcal{L}(\pi) = 0$, has been interpreted as a limit-cycle~\cite{LeePRE}. 
The presence of off-diagonal elements in $\pi$ (but not in the local states after partial tracing) is a hint of phase correlations and therefore of the presence of synchronization 
between the VdP oscillators, as can be indeed checked from the qualitative behavior of the approximated Wigner function~\cite{LeePRE, Walter2}. 
In the following we propose a quantum trajectory approach to gain a deeper look into this issue.  

The quantum trajectory formalism describes the stochastic evolution of the pure state of the system $\ket{\psi (t)}$, conditioned on measurements obtained from the continuous monitoring of the environment~\cite{milburn,trajectories}.
It has been largely used in atomic physics and quantum optics, for which the formalism was originally developed~\cite{trajectories}.
Within this approach, we can \emph{unravel} the dynamical evolution given by Eq.~\eqref{eq:master} by including the backaction of the continuous measurement process of the different environmental contributions (more details about the derivation are given in Ref.~\cite{SI}). We identify five Lindblad operators in Eq.~\eqref{eq:master}:
$L_1 = \sqrt{\gamma_\downarrow^{(1)}} a_1^2$, $L_2 = \sqrt{\gamma_\uparrow^{(1)}} a_1^\dagger$, $L_3 = \sqrt{\gamma_\downarrow^{(2)}} a_2^2$, $L_4 = \sqrt{\gamma_\uparrow^{(2)}} a_2^\dagger$, and the collective operator $L_5 = \sqrt{V} (a_1 - e^{i \theta} a_2)$ (notice that here we introduced the rates inside the definition of the Lindblad operators). The evolution can then be described by the following diffusive stochastic Schr\"odinger equation:
\begin{align} \label{eq:stochastic}
 \text{d} \ket{\psi (t)}&=\text{d}t \left[ -i H_\mathrm{eff}  + \sum_k\frac{ \langle X_k \rangle_{\psi(t)} }{2}\left(L_k - \frac{\langle X_k \rangle_{\psi(t)}}{4} \right) \right]\!\ket{\psi (t)} \nonumber \\
 &+ \sum_k \text{d}W_k(t) \left(L_k  - \frac{\langle X_k \rangle_{\psi(t)}}{2} \right)\!\ket{\psi(t)},
\end{align}
where $H_\mathrm{eff} = H - i \sum_k L_k^\dagger  L_k/2$ is a non-Hermitian (effective Hamiltonian) operator and we introduced the generalized quadrature operators $X_k = L_k + L_k^\dagger$. Here we  denoted $\langle A \rangle_{\psi(t)} \equiv \langle \psi(t)| \,A \,|\psi(t) \rangle$ as the quantum-mechanical expectation values over trajectories at time $t$.
The random variables $\text{d}W_k(t)$  are Wiener stochastic increments associated with the continuous measurement of the operators $X_k$. They follow Gaussian statistics with zero average over trajectories $\langle \text{d} W_k \rangle = 0$ and obey $\text{d}W_k^2 = \text{d}t$. The associated currents from continuous measurements read:
\begin{equation} \label{eq:currents}
 J_k(t) = \langle X_k \rangle_{\psi(t)} + \xi_k(t), 
\end{equation}
where $\xi_k(t) \equiv \text{d}W_k(t) / dt$ correspond to a white noise contribution~\cite{milburn}.

It is worth pointing out here that among different ways of unraveling the master equation dynamics \eqref{eq:master}, we choose the diffusive approach with continuous measurements of $X_k$ because it best provides information about the oscillators phases. 
Other approaches like the ones achieved by direct observation of the quantum jumps correspond to the projection of the system state in the product of local Fock basis, thereby leading to a 
randomization of the oscillators phases. Still, the persistence of signatures of synchronization in quantum jumps would be interesting to explore.

\section{Measuring synchronization}

In order to characterize synchronization between the two VdP oscillators along a single trajectory $\ket{\psi(t)}$ generated by Eq.~\eqref{eq:stochastic}, we introduce two different quantities which will help us to characterize phase-locking and synchronization of observables. The first one is the complex-valued correlator
\begin{equation}\label{eq:c}
C_\psi(t) =  \frac{\langle a_1^\dagger a_2\rangle_{\psi(t)}}{\sqrt{\langle a_1^\dagger a_1 \rangle_{\psi(t)} \langle a_2^\dagger a_2 \rangle_{\psi(t)}}},
\end{equation}
where we recall that the expectation values are taken using the stochastic wave function, $|\psi(t)\rangle$. The angle of the correlator $C_\psi \equiv |C_\psi| e^{i \Delta \phi_\psi}$ characterizes the phase difference between the two oscillators. In the classical limit, when quantum fluctuations can be neglected and the annihilation operators are replaced by the amplitudes $\alpha_i = |\alpha_i|e^{i \phi_i}$, $ C \simeq e^{i \Delta \phi}$ with $\Delta \phi = \phi_1 - \phi_2$. In general, the best quality of phase-locking $|C_\psi|\rightarrow 1$ is reached when the two oscillators are completely correlated ($|\langle a_1^\dagger a_2 \rangle| \sim \sqrt{\langle a_1^\dagger a_1 \rangle \langle a_2^\dagger a_2 \rangle}$), indicating that $\Delta \phi_\psi$ is a well defined phase. The minimum value $|C_\psi| = 0$ is instead reached when the operators are completely uncorrelated ($|\langle a_1^\dagger a_2 \rangle| = 0$) and therefore $\Delta \phi_\psi$ contains no information about the oscillators phases. 

The statistics of phase-locking along single trajectories calculated from Eq.~\eqref{eq:c} can be compared with the phase information retrieved 
from the steady state solution of the master equation $\pi$. 
From now on, we restrict ourselves to the limit $\gamma_\downarrow / \gamma_\uparrow \rightarrow \infty$ where the master equation can be analytically solved and we can 
compute the correlator $C$ in Eq.~\eqref{eq:c} (see Appendix \ref{s2}). Assuming for simplicity equal rates in both oscillators 
$\gamma_{\uparrow, \downarrow}^{(1)} = \gamma_{\uparrow, \downarrow}^{(2)} \equiv \gamma_{\uparrow, \downarrow}$ we obtain
\begin{equation} \label{eq:cpi}
 C_\pi = \frac{V (\gamma_\uparrow + V) e^{i \Delta \phi_\pi}}{(3 \gamma_\uparrow + V)\sqrt{\Delta\omega^2 + (3 \gamma_\uparrow + V)^2}},
\end{equation}
with the average phase-difference in the steady state $\Delta \phi_\pi$ defined through $\tan(\theta - \Delta \phi_\pi) = \Delta \omega/(3 \gamma_\uparrow + V)$, independent of non-linear damping. Averaging the indicator $C_\psi(t)$ in Eq.~\eqref{eq:c} for any $t$ over many trajectories, we recover $C_\pi$ in Eq.~\eqref{eq:cpi}.

A second, complementary, measure of synchronization considers 
the dynamics of local observables and the corresponding Pearson correlator \cite{Review}.
Focusing on the position quadratures $x_i = (a_i + a_i^\dagger)/\sqrt{2}$ of the two VdP oscillators, this reads:
\begin{equation}\label{eq:pearson}
 r_{x_1, x_2}(t | \Delta t) \equiv \frac{\overline{\delta \langle x_1 \rangle  \delta \langle x_2 \rangle}}{\sqrt{\overline{\delta  \langle x_1 \rangle^2} ~ \overline{\delta  \langle x_2 \rangle^2}}} 
\end{equation}
where $\delta \langle x_i \rangle \equiv \langle x_i \rangle_{\psi(t)} - 
\overline{\langle x_i \rangle}_{\psi(t)}$ and the bar stands for the 
time-average over the time-window $\Delta t$ around $t$, that is, 
$\overline{\langle x_i \rangle}_{\psi(t)} \equiv \int_{t - \Delta t/2}^{t+\Delta 
t/2} ds \langle x_i \rangle_{\psi(s)}/ \Delta t$.
The Pearson indicator takes values between $1$ and $-1$ corresponding 
respectively to perfect temporal synchronization and anti-synchronization in the 
dynamics of $\langle x_1 \rangle_{\psi(t)}$ and $\langle x_2 \rangle_{\psi(t)}$. 
For completely uncorrelated signals it becomes $0$.
It is worth noticing that the Pearson indicator \eqref{eq:pearson} does not 
capture synchronization of the positions of the two VdP oscillators in the 
average steady state dynamics as given by the density 
operator $\pi$, since 
$\langle x_i \rangle_\pi = \tr[x_i \pi] = 0$~ for 
$i=1,2$. One could consider higher moments \cite{Giorgi,Manzano}, but here we will see how quantum trajectories offer  deeper insight in the dynamical evolution
of positions, even if these vanish on average in the steady state $\pi$.

We performed numerical simulations of the two VdP oscillators system [Eq.~\eqref{eq:stochastic}] using quantum-trajectory Monte Carlo methods~\cite{qutip}. In order to investigate the steady state dynamics of the system we compute Eq.~\eqref{eq:stochastic} for pure initial states $\ket{\pi_n}$ sampled from the steady state distribution, $\pi = \sum_n \pi_n \ket{\pi_n}\bra{\pi_n}$, according to the probabilities $\pi_n$, where $\langle \pi_n | \pi_m \rangle = \delta_{n m}$. 
When averaging over measurement currents, Eq.~\eqref{eq:stochastic} reduces to the Lindblad master equation \eqref{eq:master}, where the steady state $\pi$ is recovered. 

\begin{figure}[t]
 \includegraphics[width= 1.0 \linewidth]{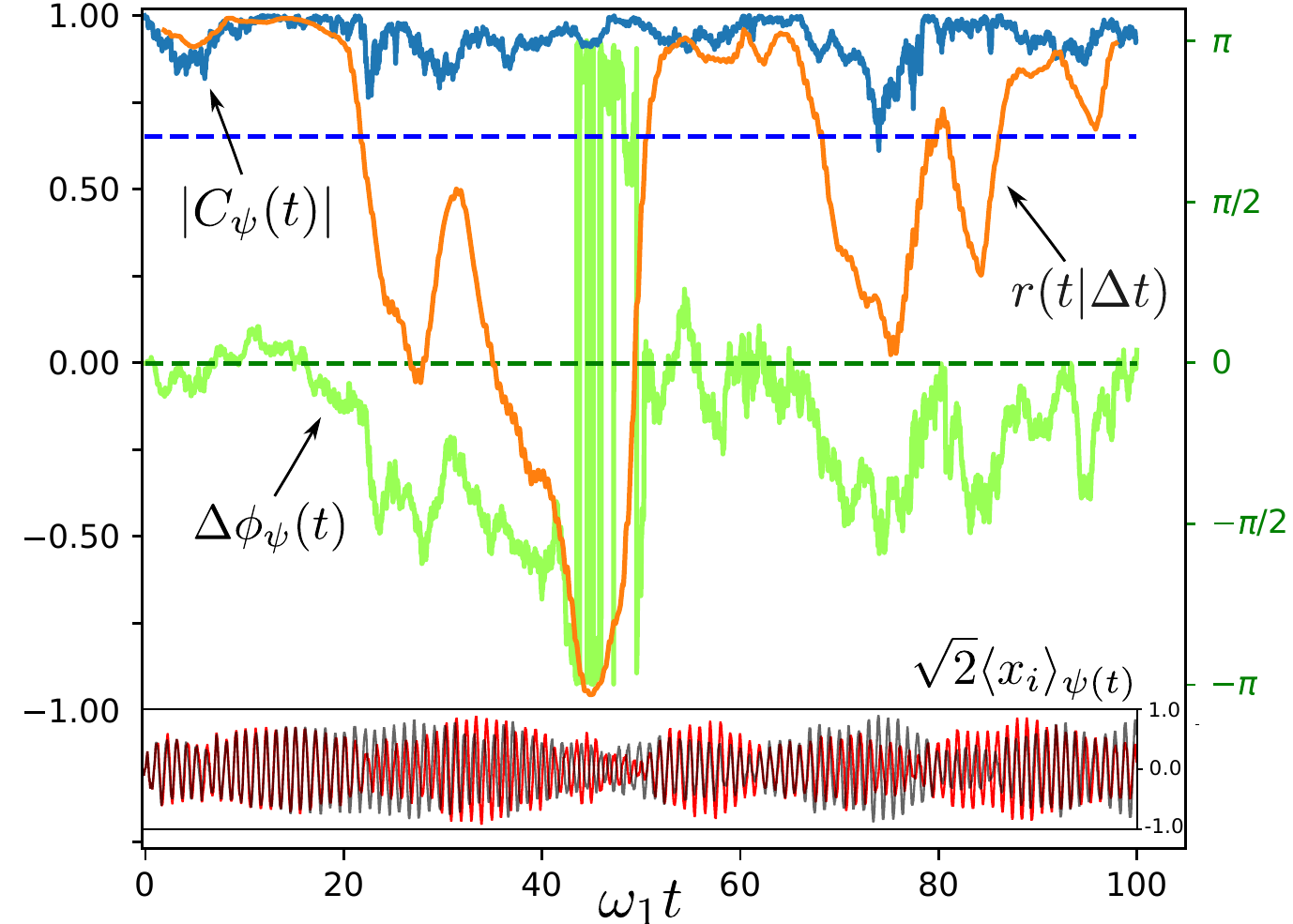}
 \caption{Modulus (blue) and phase (green) of the correlator $C(t)$ and Pearson indicator $r_{x_1, x_2}(t | \Delta t)$ (orange) as a function of time for a 
 sample trajectory $\ket{\psi (t)}$. Dashed lines correspond to average values in $\pi$. Inset: Expectation values for the positions of the two VdP oscillators $\langle x_i \rangle_{\psi(t)}$ as a function of 
 time for the same trajectory. Parameters or the simulation: $\omega_1 = 2 \pi$, $\Delta \omega = 0.1 \gamma_\uparrow$, $V= 10 \gamma_\uparrow$,
 $\gamma_\uparrow = 0.01$, $\Delta t = 8\pi /\omega_1$. }\label{fig1}
 \end{figure}

In Fig.~\ref{fig1} we show an example of the time evolution of the modulus and phase of the correlator $C_\psi(t)$ in Eq.~\eqref{eq:c} as well as the Pearson indicator $r_{x_1, x_2}(t|\Delta t)$ over a single trajectory $\ket{\psi(t)}$ as a function of time. 
We focus on the transition regime to phase locking. The corresponding average values obtained from Eq.~\eqref{eq:cpi} are respectively the top and bottom dashed lines. 
The Pearson indicator changes during the evolution and drops down whenever the phase difference departs from $\theta = 0$, consistently with the local observables on the oscillators trajectories, $\langle x_i \rangle_{\psi(t)}$ ($i= 1, 2$).
Indeed the relative phase is not locked to a fixed value, displaying instead a slow time dependence,  $\Delta \phi_\psi (t)$, which can highly depart from its average value $\Delta \phi_\pi = - 0.008$ (bottom dashed line). Still, the modulus $|C(t)|$ shows a significant correlation in the trajectories of the VdP oscillators during this time interval, even if the average value is moderate (upper dashed line), indicating a high accuracy of the phase-difference between the oscillators. This means that trajectories which are actually not phase-locked to $\theta$ may instead contribute with a high value to $|C_\pi|$, giving the (wrong) impression that the system is synchronized, and therefore spotting the necessity for looking at synchronization indicators beyond average values.

The present approach also enables us to explore the relation between the emergence of synchronization and the entanglement shared between the two VdP oscillators during single trajectories, as first considered in Refs. \cite{Nha,Viviescas}. 
The quantum state of the two oscillators remains pure during the whole trajectory [Eq.~\eqref{eq:stochastic}] due to the incorporation of the environmental measured currents $J_k(t)$ in Eq.~\eqref{eq:currents}. Therefore the entanglement entropy is a unique measure of entanglement~\cite{Ent1, Ent2}, namely
\begin{equation} \label{eq:ent}
 S_{\psi}(t) = - \tr_1[\rho_\psi(t) \log \rho_\psi(t)],
\end{equation}
with the reduced state of oscillator 1 during a stochastic trajectory $\rho_{\psi}(t)= \tr_2[\ket{\psi(t)}\bra{\psi(t)}]$, and where 
we denote by $\tr_i$ the partial trace with respect to degrees of freedom of oscillator $i$. The entanglement entropy $S_{\psi}(t)$ for the VdP oscillators in the quantum regime takes values between $S_{\psi} = 0$ (no entanglement) and $S_{\psi}=\log 2 \sim 0.69$ (maximally entangled state). The average of $S_{\psi}(t)$ among trajectories defines the average entanglement associated with the environmental monitoring scheme~\cite{Nha, Viviescas}. It is lower bounded by (but it does not necessarily correspond to) the entanglement  of formation in the (mixed) steady state $\pi$, i.e. the minimum average entanglement of any possible decomposition of $\pi$ into pure states. Nevertheless, in the following we will focus more on higher order moments of the $S_\psi$ distribution rather than on its average.

\section{STATISTICS OF SYNCHRONIZATION} \label{sec:simu}

\begin{figure}[t]
 \includegraphics[width= 1.0 \linewidth]{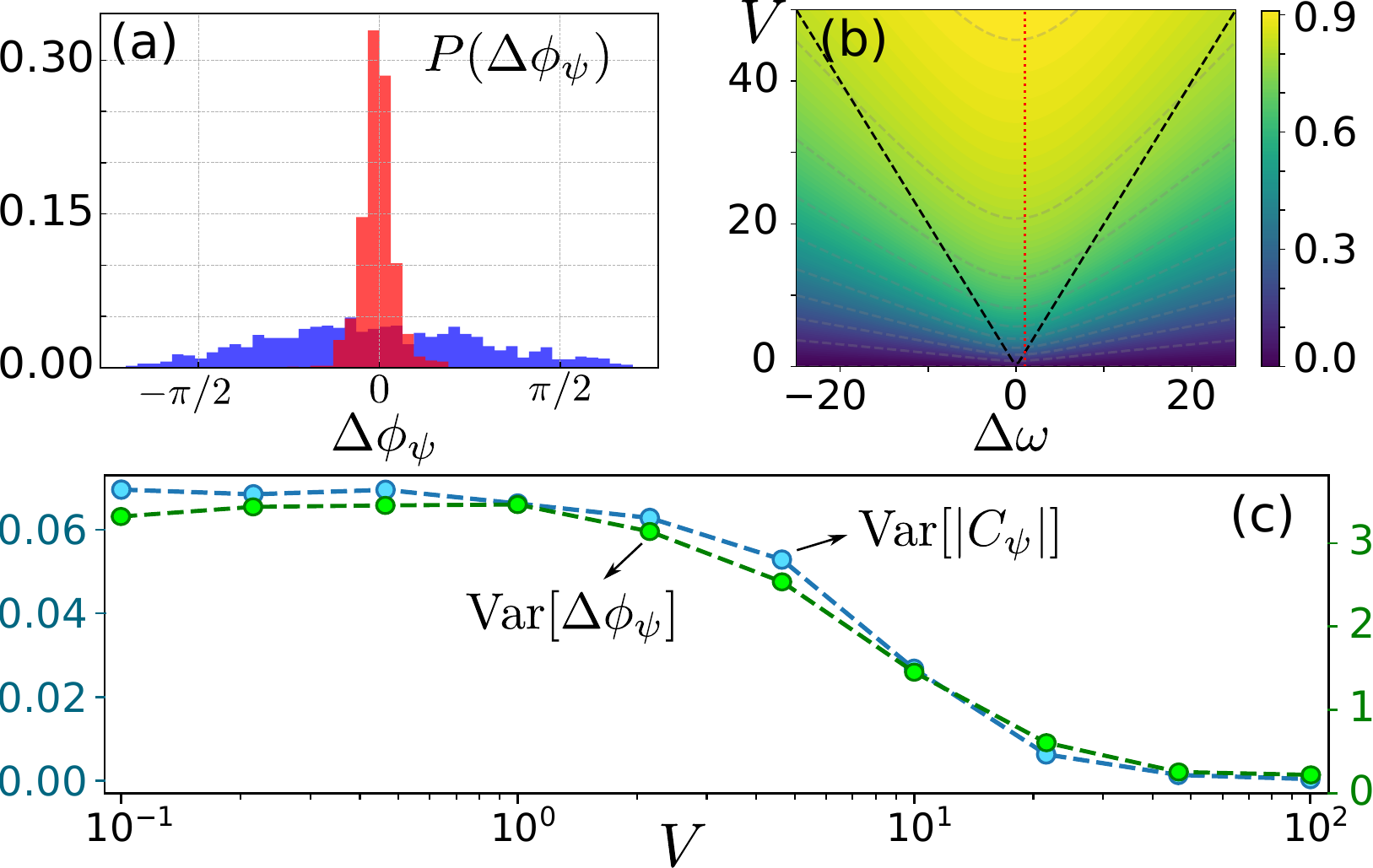}
 \caption{ (a) Probability distribution of the time-averaged phase-difference along trajectories $\Delta \phi_\psi$ for $\Delta \omega = \gamma_\uparrow$ and two different choices of the coupling strength $V = \{ 5 \gamma_\uparrow, 50 \gamma_\uparrow \}$ (blue and red bars, respectively). (b)  Classical Arnold tongue (dashed black lines) and modulus of the steady-state correlator $|C_\pi|$. (c) Variances of the distributions $P(|C_\psi|)$ and $P(\Delta \phi_\psi)$ as a function of the coupling strength $V$ for same detuning. Detuning $\Delta \omega$ and $V$ are plotted in units of $\gamma_\uparrow$. Other parameters: $\hbar \omega_1 = 8\pi$, $\gamma_\uparrow = 0.01$, and $10^3$ trajectories.} \label{fig2}
 \end{figure}

Evaluating the different measurements of phase-locking and synchronization of observables reported above along many trajectories, we are able to numerically reconstruct the full probability distributions of the measures $|C_\psi(t)|$, $\Delta \phi_\psi(t)$, $r_{x_1, x_2}(t|\Delta t)$ and $S_\psi(t)$ at any given instant of time $t$. We denote these probability densities by  $P_t(|C_\psi|)$, $P_t(\Delta \phi)$, $P_t(r_{x_1,x_2})$ and $P_t(S_\psi)$ respectively. Nevertheless, since the trajectories are computed for the steady-state dynamics, these probability distributions are, up to finite-size sampling errors, independent of time. Therefore, in order to reduce statistical errors, we compute their time-averaged versions from some initial time $t>0$ until a final fixed time, such that $\ket{\psi(t)}$ has sufficient time to depart from the initial state $\ket{\pi_n}$ sampled from $\pi$. We refer to the time-averaged probability distributions as $P(|C_\psi|)$, $P(\Delta \phi_\psi)$, $P(r_{x_1,x_2})$ and $P(S_\psi)$.

\begin{figure}[t]
 \includegraphics[width= 1.0 \linewidth]{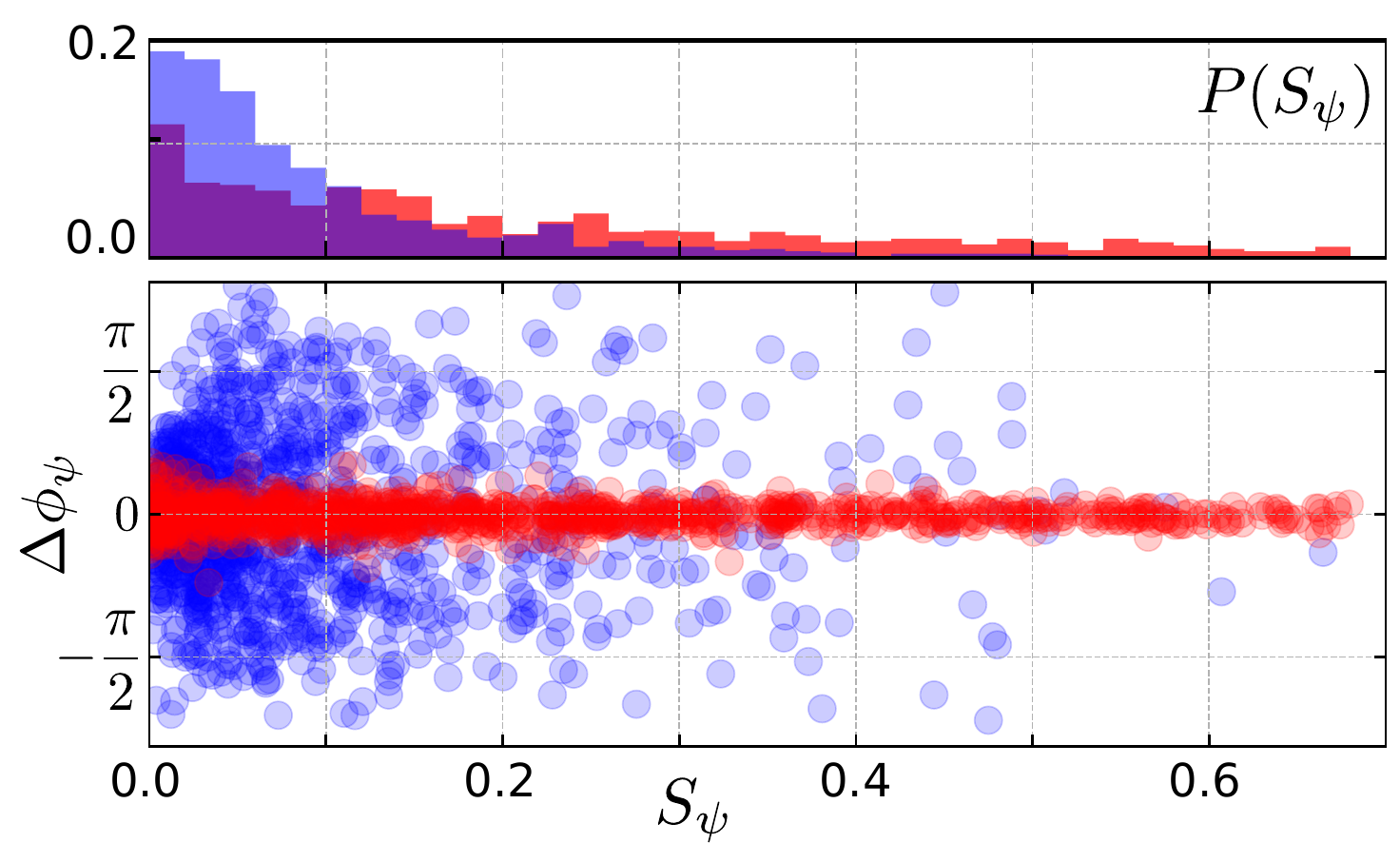}
 \caption{Scatter plot for the entanglement $S_\psi$ shared by the VdP oscillators during trajectories and their phase-difference $\Delta \phi_\psi$ 
 for two different values of the coupling strength $V = \{ 5 \gamma_\uparrow, 50 \gamma_\uparrow \}$ (blue circles and red circles respectively). 
 Top inset: Entanglement probability distributions $P(S_\psi)$ for the two cases. Other parameters are as in Fig. \ref{fig2}.} \label{fig3}
 \end{figure}

\begin{figure}[t]
    \includegraphics[width= 1.0 \linewidth]{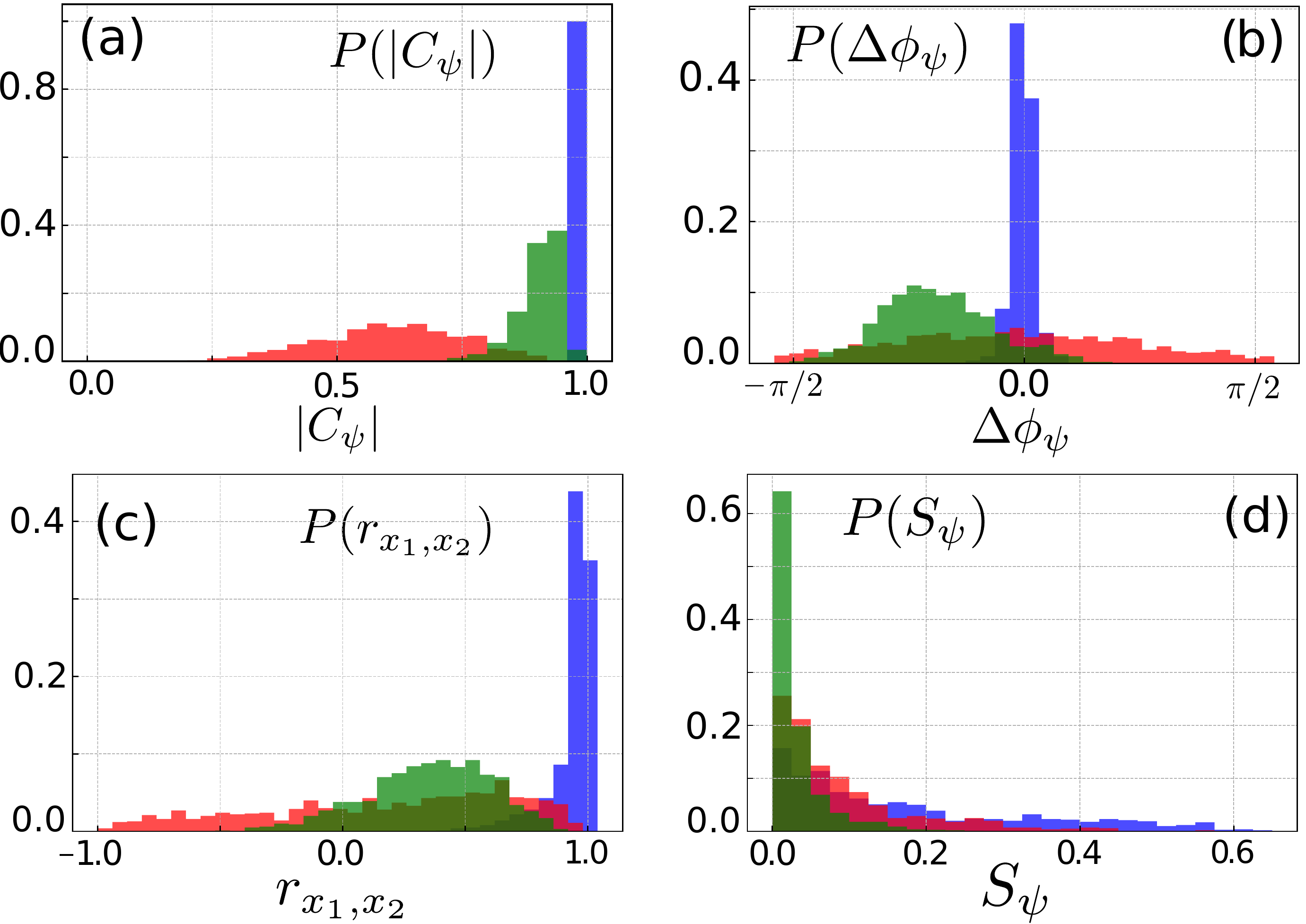}
    \caption{Time-averaged probability distributions of (a)$ \mid C_\psi \mid$, (b) phase-differences $\Delta \phi_\psi$, (c) the Pearson indicator $r_{x_1, x_2}$ and (d) the entanglement entropy $S_\psi$, for $10^{3}$ trajectories for three cases: $\Delta \omega = 1.0 \gamma_{\uparrow}$ and $V = 100 \gamma_{\uparrow}$ (blue bars), $\Delta\omega = 1.0 \gamma_\uparrow$ and $V = 5 \gamma_{\uparrow}$ (red bars), and $\Delta \omega = 20 \gamma_\uparrow$ and $V = 20 \gamma_{\uparrow}$ (green bars).} 
    \label{fig:s1}
\end{figure}

\begin{figure}[t]
 \includegraphics[width= 0.85 \linewidth]{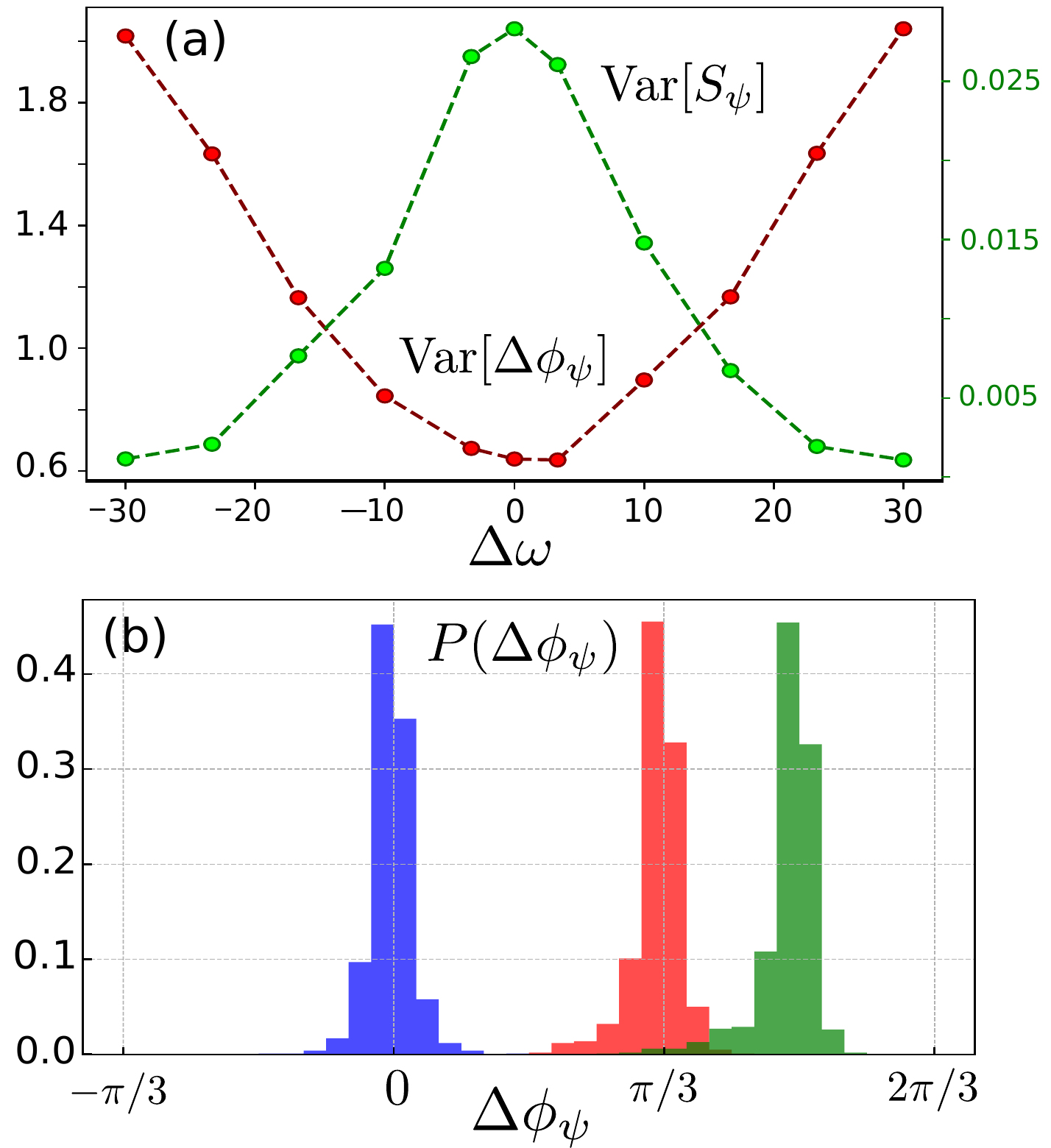}
  \caption{(a) Variance of the time-averaged phase difference, $\Delta \phi_\psi$, and the entanglement entropy, $S_\psi$, for 11 values of detuning $\Delta\omega \in [-30 \gamma_\uparrow, 30\gamma_\uparrow]$ and fixed $V= 20 \gamma_\uparrow$. (b) Probability distributions of the phase-difference, $P(\Delta \phi_\psi)$, for three values of the phase-locking angle $\theta = \{ 0, \pi/3, \pi/2 \}$ (blue, red, green) with $\Delta \omega = \gamma_\uparrow$ and $V= 100 \gamma_\uparrow$. Other parameters are the same as those in Fig.~\ref{fig:s1}(a).} 
  \label{fig:s2}
\end{figure}

We find that phase-locking in the model can be detected and characterized from the shape of the probability distributions 
$P(|C_\psi|)$ and $P(\Delta \phi_\psi)$ (Fig. \ref{fig2}) and $P(r_{x_1,x_2})$. Moreover, comparing these distributions with the entanglement probability distribution, $P(S_\psi)$, a persistent relation between synchronization and entanglement along trajectories is observed. Synchronized trajectories tend to share a greater amount of entanglement than unsynchronized ones. That is, trajectories show high values of entanglement more often when we approach high-quality synchronization regimes (Fig.~\ref{fig3}). 
The trends observed for the probability distributions of the different synchronization measures allowing the characterization of synchronization are robust, and observed for a broad range of parameters in the model (Figs. \ref{fig:s1} and \ref{fig:s2}).

In Fig. \ref{fig2}(a) we show two different instances of the phase-differences 
probability distribution, $P(\Delta \phi_\psi)$, for a fixed detuning between oscillators $\Delta \omega = \gamma_\uparrow$ and two different choices of the 
coupling strength $V = \{ 5 \gamma_\uparrow, 50 \gamma_\uparrow \}$. In Fig 2(b) we show the classical Arnold tongue (region inside the black dashed lines) 
together with a color map displaying $|C_\pi|$ in Eq.~\eqref{eq:cpi}. In Fig.~\ref{fig2}(c) we plot the variance of the distributions $P(\Delta \phi_\psi)$ 
and $P(|C_\psi|)$ as a function of $V$ for same detuning. We see that for values inside the (classical) Arnold tongue, small values of $V$ induce a phase-differences 
distribution smoothly peaked at $\Delta \phi_\psi = \theta = 0$ with a large variance. If $V$ is increased $P(\Delta \phi_\psi)$ becomes sharp around 
$\Delta \phi_\psi = 0$ and both variances $\mathrm{Var}[\Delta \phi_\psi]$ and $\mathrm{Var}[|C_\psi|]$ approach zero. The red dotted line in Fig. \ref{fig2}(b) correspond to the parameters used in Fig. 2(a). As can be appreciated in both Figs.~\ref{fig2}(a) 
and Figs.~\ref{fig2}(b), for small values of the coupling $V$ a very poor phase-locking is expected even for small detunings $\Delta \omega \rightarrow 0$. This is in contrast to the classical case, which predicts phase-locking inside all the region.

Figure \ref{fig3} shows the statistical correlations between synchronization and entanglement during single trajectories for synchronized (red circles) and unsynchronized (blue circles) regimes. Each point represents a single trajectory for which we computed (the time-averages of) $\Delta \phi_\psi$ and $S_\psi$. In the top inset, the corresponding entanglement probability distributions $P(S_\psi)$ reconstructed from the data are shown. We see that inside the good synchronization region, for $V= 50 \gamma_\uparrow$ (red circles), phase-locked trajectories show high values of entanglement (arriving up to the maximal value $S_\psi \simeq \log 2$) more frequently, as manifested in the long tail of the (red) probability distribution $P(S_\psi)$. Instead, when synchronization is poor, $V= 5 \gamma_\uparrow$ (blue circles), this effect tends to disappear and no correlation between phase and entanglement can be inferred from the data. In this case the tail in the (blue) entanglement probability distribution is lost. This statistical correlation for the tails of the distribution provides a new link between a purely dynamical phenomenon, namely, synchronization (and in particular phase-locking) with a strong measure of quantum correlations, entanglement, along trajectories. 
 
The shape of the probability distributions $P(\Delta \phi_\psi)$, $P(|C_\psi|)$, $P(r_{x_1, x_2})$ and $P(S_\psi)$ can be better appreciated in Fig.~\ref{fig:s1} where the four full probability distributions are shown for a sample of $10^3$ trajectories in the regimes of nearly perfect synchronization (blue bars,  $V=100\gamma_{\uparrow}$ and $\Delta\omega= \gamma_{\uparrow}$), poor synchronization inside the classical Arnold tongue (red bars, $V=5\gamma_{\uparrow}$ and $\Delta\omega= \gamma_{\uparrow}$), and poor synchronization outside the classical Arnold tongue (green bars, $V=20\gamma_{\uparrow}$ and $\Delta\omega=20\gamma_{\uparrow}$). In the regime of the perfect in-phase synchronization (blue bars in Fig.~\ref{fig:s1}) the distribution $P(|C_\psi|)$ is highly peaked around its average value $C_\pi \simeq 0.99$, the distribution of phase-differences, $P(\Delta \phi_\psi)$, is peaked around $\theta = 0$, and the distribution of the Pearson indicator, $P(r_{x_1, x_2})$ is peaked around the maximum value $r_{x_1, x_2} \simeq 1$. This is accompanied by a large tail in the probability distribution of entanglement $P(S_\psi)$. On the contrary, in the other two cases (red and green bars in  Fig.~\ref{fig:s1}) all the distributions for the synchronization indicators become much more flattened, spreading along all their ranges. This is a signature of a poor synchronization, even if the average values may differ in the two cases. Also in both cases $P(S_\psi)$ becomes sharp around $0$, meaning that entanglement is not produced in almost all trajectories. 
 
In Fig.~\ref{fig:s2}(a) we plot the variance of the distribution $P(\Delta \phi_\psi)$ (red line) as a function of $\Delta \omega$ for a fixed value of the dissipative coupling strength $V=20\gamma_{\uparrow}$. There we can see how, despite that we are still in a regime of moderate-bad synchronization, the later improves when $\Delta \omega \rightarrow 0$ as expected from the classical case, since the variance of the distribution becomes small. Comparing with the variance of the entanglement probability distribution $P(S_\psi)$ (green line), we see that it behaves in the opposite way. That is, the variance of $P(S_\psi)$ increases whenever synchronization becomes stronger, in line with the appearance of long tails in the entanglement distribution reported above. This means that the probability to see a trajectory with a high value of entanglement becomes greater when $\Delta \omega$ is close to zero, that is, when we enter the region of parameters where many trajectories show good synchronization.

Finally, in Fig.~\ref{fig:s2}(b) we provide an additional plot showing $P(\Delta \phi_\psi)$ when varying the phase-locking angle $\theta$ introduced in Eq.~\eqref{eq:master}. We focus on parameters leading to good synchronization ($V=100 \gamma_\uparrow$ and $\Delta \omega = \gamma_\uparrow$) to show the existence of phase-locking at the input angle $\theta$ also when it takes other values different from zero. Blue bars correspond to $\theta = 0$, red bars are for $\theta = \pi/3$, and green ones stand for $\theta = \pi/2$. As can be seen, phase-locking is verified at the different angles $\theta$ with the probability distributions showing analogous features than for the case $\theta = 0$.

\section{Discussion and Conclusions}

We have shown that synchronization can arise in quantum trajectories, here for quantum Van der Pol oscillators with dissipative coupling, providing deeper insights about the synchronization phenomenon in the quantum regime. 
Departures from the classical scenario are reported in the limit of almost identical weakly coupled oscillators.
The monitored system also displays a clear connection between synchronization entailed by phase-locking and entanglement 
in quantum trajectories, spotted by the emergence of long tails in the entanglement distribution. 
This phenomenon is compatible with previous results showing a link between the entanglement of formation of the steady state $\pi$ and synchronization in a region reminiscent of the Arnold tongue~\cite{LeePRE}. We actually find small values for the variance of $P(\Delta \phi_\psi)$ accompanied by high values for the variance of $P(S_\psi)$ for similar parameters [see e.g. Fig.~\ref{fig:s2}(a)]. However, the suppression of the large tails in $P(S_\psi)$ when decreasing $V$ inside the classical Arnold tongue region is smooth [red bars in Fig.~\ref{fig:s1}(d)],  in contrast to the entanglement of formation of $\pi$, which becomes suddenly zero for $V \sim 10 \gamma_\uparrow$~\cite{LeePRE}. A deeper comparison with the entanglement of formation in $\pi$ and its reconstruction beyond the quantum limit, may be performed by extending the optimal diffusive unraveling for entanglement detection proposed in Ref.~\cite{Viviescas} to the present situation.

It would be also interesting to explore connections and possible applications to quantum control~\cite{carmichael, dicarlo, haroche}, 
quantum information processing~\cite{barreiro, Davide, Jordan} or quantum thermodynamics along trajectories~\cite{romito,gong,naghiloo, martingales}. Other possible extensions of this work include considering reactive instead of dissipative couplings as well as other systems amenable to experimental realizations, such as optomechanical systems, atomic systems or superconducting qubits. 
In this context, it may be relevant to extend our results to the case of imperfect detection schemes, where the finite efficiency of the detectors or the impossibility to unravel some of the Lindblad operators, leads to a description in terms of a stochastic master equation~\cite{Jacobs,milburn}. In such a case, the synchronization indicators introduced here could be directly applied to the corresponding stochastic density operator, but more attention should be paid to the choice of a reliable measure of entanglement along single trajectories, since it will now require a minimization procedure.

To sum up, we believe that the approach introduced in the present paper opens new possibilities of more precise determination of synchronization in the quantum regime by looking at the statistical properties of different synchronization indicators. Importantly, this approach also helps to unveil a hidden link between synchronization along single trajectories and the generation of entanglement. In addition, our results offer an operationally well-defined way to experimentally characterize quantum synchronization in systems where environmental monitoring becomes possible.

\begin{acknowledgments}
 R.Z. acknowledges support from MINECO/AEI /FEDER through projects EPheQuCS FIS2016-78010-P and the Mar\'ia de Maeztu Program for Units of Excellence in R$\&$D (MDM-2017-0711).
\end{acknowledgments}

\appendix

\section{VdP steady state} \label{s2}
In this appendix we analytically obtain the steady-state density matrix $\pi$ of the two VdP oscillators from the master equation \eqref{eq:master} in the limit $\gamma_{\downarrow}/\gamma_{\uparrow}\longrightarrow\infty$. 
For simplicity we also assume symmetric rates in both oscillators, that is, $\gamma_{\downarrow,\uparrow}^{(1)}=\gamma_{\downarrow,\uparrow}^{(2)}=\gamma_{\downarrow,\uparrow}$.
In this limit, the VdP oscillators are restricted to their two lowest Fock states, $\arrowvert 0 \rangle_i$ and $\arrowvert 1 \rangle_i$ since any other state is annihilated by the non-linear damping term in Eq.~(1)~\cite{LeePRL}. 
This implies that the master equation can be mapped to a dissipative spin model of the form~\cite{LeePRE}:
\begin{align} \label{eq:spin}
\dot{\rho} = ~\mathcal{L}(\rho) &= -i[H,\rho] + V \mathcal{D}[\sigma_1^{-} - e^{i \theta} \sigma_2^{-}]\rho \nonumber \\ 
&+ \sum_{j=1}^2 2 \gamma_\uparrow^{(j)} \mathcal{D}[\sigma_j^{-}]\rho + \gamma_\uparrow^{(j)} \mathcal{D}[\sigma_j^{+}]\rho.
\end{align}
Here the Hamiltonian reduces to ${H}= \sum_{j=1,2}\hbar \omega_{j}{\sigma}_{j}^{+}{\sigma}_{j}^{-}$, and the oscillator ladder operators $a$ and $a^\dagger$ are transformed in spin-flip operators 
${\sigma}_{j}^{-}=\vert 0\rangle \langle 1\vert_j$ and ${\sigma}_{j}^{+}=\vert 1\rangle \langle 0\vert_j$. Importantly, in Eq.~\eqref{eq:spin} the original non-linear damping term appearing in \eqref{eq:stochastic}, has been replaced by a linear damping with an effective rate $2 \gamma_{\uparrow}$. This can be understood from the fact that any transition $\ket{1} \rightarrow \ket{2}$ in the original model promoted by the pumping term (at a rate $2 \gamma_\uparrow$), will immediately decay to $\ket{2}\rightarrow \ket{0}$ as $\gamma_\downarrow \rightarrow \infty$, leading to an effective transition $\ket{1} \rightarrow \ket{0}$.
 
Following Ref.~\cite{LeePRE}, the steady state solution of Eq.~\eqref{eq:spin} is obtained from $\mathcal{L}(\pi)=0$, whose non-zero elements read:
\begin{align} \label{eq:ss}
\langle00\vert\pi\vert00\rangle &= 1-\frac{\gamma_{\uparrow}(5\gamma_{\uparrow}+2V)[\Delta\omega^{2}+(3\gamma_{\uparrow}+V)^{2}]}{N}, \\
\langle01\vert\pi\vert01\rangle &=\frac{\gamma_{\uparrow}(2\gamma_{\uparrow}+V)[\Delta\omega^{2}+(3\gamma_{\uparrow}+V)^{2}])}{N}, \\
\langle11\vert\pi\vert11\rangle &= \frac{\gamma_{\uparrow}^{2}[\Delta\omega^{2}+(3\gamma_{\uparrow}+V)^{2}])}{N}, \\
\langle01\vert\pi\vert10\rangle &=\frac{\gamma_{\uparrow}V(\gamma_{\uparrow}+V)(3\gamma_{\uparrow}+V-i\Delta\omega)e^{-i\theta}}{N}, \label{eq:ss4}
\end{align}
and we have $\langle10\vert\pi\vert10\rangle = \langle01\vert\pi\vert01\rangle$, and $\langle10\vert \pi \vert01\rangle^{*}=\langle01\vert\pi\vert10\rangle$. Here we introduced $N=(3\gamma_{\uparrow}+V)[3\gamma_{\uparrow}(\Delta\omega^{2}+9\gamma_{\uparrow}^{2})+(\Delta\omega^{2}+27\gamma_{\uparrow}^{2})V+8\gamma_{\uparrow}V^{2}]$.

The marginal states of the two VdP oscillators can be computed by partial tracing $\pi$ over the complementary oscillator, $\pi_i \equiv \tr_j[\pi] = (1 - p)\ket{0}\bra{0} + p \ket{1}\bra{1}$, where $i \neq j$ and
\begin{equation} \label{eq:p}
p \equiv \gamma_\uparrow (3\gamma_\uparrow + V)) [\Delta \omega^2 + (3\gamma_\uparrow + V)^2]/N.
\end{equation}
The marginal states $\pi_i$ have free-phase (no off-diagonal elements) and a population ratio between ground and excited states given by $1-p:p$. We note from Eq.~\eqref{eq:p} that $p$ is strictly greater than zero whenever $\gamma_\uparrow$ is finite. Therefore the oscillators never collapse to their ground states. We also have $p \leq 1/3$, the maximum being reached in the limit of uncoupled oscillators, $V \rightarrow 0$, where the population ratio becomes $2:1$. Increasing $\Delta \omega$ and $V$ we obtain lower values of $p$, and in the limit $V \gg \gamma_\uparrow$ we have $p \rightarrow 1/8$ independently of the detuning $\Delta \omega$. The presence of larger drops in $p$ outside the classical Arnold's tongue region for moderate values of $V > \gamma_\uparrow$ can be seen as a manifestation of the classical phenomenon of oscillations collapse in the quantum regime as discussed in Ref.~\cite{Kanamoto}.

Using Eqs.~\eqref{eq:ss}-\eqref{eq:ss4} we can now calculate the value of the complex-value correlator $C$ introduced in Eq.~\eqref{eq:cpi} for the steady state $\pi$. We obtain:
\begin{align}
C_{\pi}&=\frac{\langle \hat{\sigma}_{1}^{+}\hat{\sigma}_{2}^{-}\rangle}{\sqrt{\langle \hat{\sigma}_{1}^{+}\hat{\sigma}_{1}^{-}\rangle\langle \hat{\sigma}_{2}^{+}\hat{\sigma}_{2}^{-}\rangle}} \nonumber \\
&= \frac{V(\gamma_{\uparrow}+V)}{(3\gamma_{\uparrow}+V)\sqrt{\Delta\omega^{2}+(3\gamma_{\uparrow}+V)^{2}}} ~e^{i \Delta \phi_{\pi}}, 
\end{align}
where $\Delta\phi_{\pi}$ is the phase difference of the two coupled VdP oscillators defined through:
\begin{equation} \label{eq4}
\tan(\theta - \Delta\phi_{\pi}) = \frac{\Delta\omega}{3\gamma_{\uparrow}+V}.
\end{equation}
In Fig.\ref{fig2}(b) of Sec.\ref{sec:simu}, we plot $|C_\pi|$ as a function of $V$ and $\Delta \omega$ and compare to the Classical Arnold tongue. 
As can be seen there an important region inside the Arnold tongue corresponding to small values of the detuning $\Delta \omega$ and small values of $V$ (as compared to $\gamma_\uparrow$) $|C_\pi|$ can be far from $1$. This implies a smooth transition from no-synchronized to synchronized regimes. In the transition regime, the average phase difference between the oscillators, $\Delta \phi_\pi$, may therefore be poorly informative due to the presence of quantum fluctuations.

\clearpage

\onecolumngrid

\widetext
\begin{center}
\textbf{\large Supplemental Material: Synchronization along Quantum Trajectories}
\end{center}

\setcounter{equation}{0}
\setcounter{figure}{0}
\setcounter{table}{0}
\makeatletter
\renewcommand{\thesection}{S\arabic{section}}
\renewcommand{\theequation}{S\arabic{equation}}
\renewcommand{\thefigure}{S\arabic{figure}}
\renewcommand{\bibnumfmt}[1]{[S#1]}
\renewcommand{\citenumfont}[1]{S#1}

\

In this Supplemental Material we provide more technical details on the dynamical evolution of the two Van der Pol oscillators under environmental monitoring. In particular we include a derivation of the diffusive stochastic Schr\"ondiger equation employed in the main text. 

\section*{Diffusive stochastic Schr\"odinger equation} \label{s1}

We show how to obtain the diffusive stochastic Schr\"odinger equation (2) in the main text following the derivations in Refs.~\cite{Wiseman, Wiseman93, Manzanothesis}. Our starting point is the Lindblad master equation (1) in the main text, which we will unravel by using a generalized homodyne detection scheme.
For convenience we will rewrite (1) as:
\begin{equation} \label{eq:s1}
 \dot{\rho} = ~\mathcal{L}(\rho) = -i[H,\rho] + \sum_k L_k \rho L_k^\dagger - \frac{1}{2}\{L_k^\dagger L_k, \rho \},
\end{equation}
for the Lindblad operators $L_1 = \sqrt{\gamma_\downarrow^{(1)}} a_1^2$, $L_2 = \sqrt{\gamma_\uparrow^{(1)}} a_1^\dagger$, $L_3 = \sqrt{\gamma_\downarrow^{(2)}} a_2^2$, $L_4 = \sqrt{\gamma_\uparrow^{(2)}} a_2^\dagger$, and the collective operator $L_5 = \sqrt{V} (a_1 - e^{i \theta} a_2)$, which include the corresponding rates.

We notice here the following Gauge symmetry of (2), for which a double transformation $L_k \rightarrow L_k^\prime = L_k + l_k$ and $H \rightarrow H^\prime = H - i\sum_k (L_k l_k^\ast + L_k^\dagger l_k)/2$ leaves invariant Eq.~\eqref{eq:s1}.
Therefore we substitute $L_k$ and $H$ by $L_k^\prime$ and $H^\prime$ in Eq.~\eqref{eq:s1} and unravel it using the standard direct detection scheme. When the reservoir is assumed to be made of harmonic modes, 
like electromagnetic radiation, adding the displacement $l_k$ to the Lindblad operators corresponds to the implementation of Homodyne detection schemes~\cite{Wiseman}. Here we apply the same unraveling methods in a generic situation 
having in mind the same physical interpretation as in the Homodyne measurement of field-quadratures~\cite{Wiseman93}.

The evolution is split in an infinite sequence of intervals of infinitesimal duration $dt$, where the dynamics is updated according to a completely positive and trace preserving (CPTP) map $\rho_{t + dt} = \mathcal{E}(\rho_t) = \sum_n M_n(dt) \rho_t M_n^\dagger(dt)$ with Kraus operators:
\begin{align} \label{eq:jumps}
 M_0(dt) &= \id - dt\left(iH + \frac{1}{2} \sum_k L_k^{\prime \dagger} L_k^\prime \right) =  \id - dt\left(iH  + \frac{1}{2} \sum_k L_k^{\dagger} L_k + |l_k| X_k + |l_k|^2 \right), \\
 M_k(dt) &= \sqrt{dt} L_k^\prime = \sqrt{dt}(L_k + l_k),
\end{align}
with $X_k = L_k e^{-i\varphi_k} + L_k^\dagger e^{i\varphi_k}$, and $l_k = |l_k| e^{i \varphi_k}$. Here the operators $M_k$ correspond to the detection of a jump of type $L_k^\prime$ in the dynamical evolution, while the operator $M_0$ stand for the intervals where no jumps of any type are detected.
Assuming that at time $t$ the state of the system is the pure state $\ket{\psi(t)}$, their probabilities read
\begin{align}\label{eq:probs}
 P_0(dt) = 1 - dt \sum_k \langle L_k^\dagger L_k + |l_k| X_k + |l_k|^2 \rangle_{\psi(t)}, \\
 P_k(dt) = dt \sum_k \langle L_k^\dagger L_k + |l_k| X_k + |l_k|^2 \rangle_{\psi(t)},
\end{align}
where $\langle A \rangle_{\psi(t)} \equiv \bra{\psi(t)} A \ket{\psi(t)}$ is the expectation value along the trajectory at time $t$. It can be easily verified that $P_0(dt) + \sum_k P_k(dt) = 1$.

As can be readily appreciated from Eqs.~\eqref{eq:probs}, whenever $|l_k|$ is  order $1$, the probability of having any jump $L_k^\prime$ is only of order $dt$, while the probability of having no jumps during the interval $dt$ is of order $1$. 
Therefore the different type of jumps correspond to Poisson processes, almost all the time no jumps of type $L_k^\prime$ will be detected, and the evolution of the system will occur according to the operator $M_0(dt)$. That is:
\begin{equation}
 \ket{\psi^{(0)}(t + dt)} = \frac{M_0}{\sqrt{P_0(dt)}} \ket{\psi(t)} = \ket{\psi(t)} - dt \left(i H + \frac{1}{2} \sum_k \left(L_k^{\dagger} L_k  - \langle L_k^{\dagger}L_k \rangle_{\psi(t)} \right) + \frac{1}{2} \sum_k |l_k|\left(X_k - \langle X_k \rangle_{\psi(t)}\right) \right) \ket{\psi(t)}, 
\end{equation}
which corresponds to a smooth non-unitary evolution. On the other hand, at some (rare) instant of times, where a jump $k$ is detected, the system state changes as:
\begin{equation}
\ket{\psi^{(k)}(t + dt)} = \frac{M_k}{\sqrt{P_k(dt)}} \ket{\psi(t)} = \sqrt{dt}~ \frac{L_k + l_k}{\sqrt{P_k(dt)}} \ket{\psi(t)} = \frac{L_k + l_k}{\sqrt{\langle (L_k^\dagger + l_k^\ast)(L_k + l_k) \rangle_{\psi(t)}}} \ket{\psi(t)}.
\end{equation}

The stochastic Schr\"odinger equation can be constructed by introducing the number of jumps of each type $k$ detected until time $t$, $N_k(t)$. Whenever the probabilities $P_k(dt)$ remain of order $dt$ the number of jumps fulfill Poisson statistics and 
the associated stochastic increments $dN_k(t)$ fulfill $dN_k(t) dN_l(t) = \delta_{k l} dN_k(t)$, with average over trajectories $\langle dN_k(t) \rangle = P_k(dt)$. The quantities $dN_k(t)$ are stochastic variables taking values either $0$ (when no jumps are detected) 
or $1$ when a jump $k$ is detected. The infinitesimal time-evolution of the system $d\ket{\psi(t)} \equiv \ket{\psi(t+dt)} - \ket{\psi(t)}$ can then be written in It$\hat{\mathrm{o}}$ form as a sum of the different pieces of the evolution:
\begin{align}
 d \ket{\psi(t)} &= dt\left[1- \sum_k dN_k(t)\right]\left(- iH - \frac{1}{2} \sum_k \left(L_k^{\dagger} L_k  - \langle L_k^{\dagger} L_k \rangle_{\psi(t)}\right) - \frac{1}{2} \sum_k |l_k|\left(X_k - \langle X_k \rangle_{\psi(t)}\right) \right) \ket{\psi(t)} \nonumber \\
 &~~+ \sum_k dN_k(t)\left( \frac{L_k + l_k}{\sqrt{\langle (L_k^\dagger + l_k^\ast)(L_k + l_k) \rangle_{\psi(t)}}} - \id \right) \ket{\psi(t)},
\end{align}
which, by noticing that $dt dN_k(t) \sim O(dt^2)$, leads to the standard form of the stochastic Schr\"odinger equation for jumps $L_k^\prime = L_k + l_k$:
\begin{align}\label{eq:sto}
 d \ket{\psi(t)} &= dt \left(- iH - \frac{1}{2} \sum_k \left(L_k^{\dagger} L_k  - \langle L_k^{\dagger} L_k \rangle_{\psi(t)}\right) - \frac{1}{2} \sum_k |l_k|\left(X_k - \langle X_k \rangle_{\psi(t)}\right) \right) \ket{\psi(t)} \nonumber \\
 &~~+ \sum_k dN_k(t)\left( \frac{L_k + l_k}{\sqrt{\langle (L_k^\dagger + l_k^\ast)(L_k + l_k) \rangle_{\psi(t)}}} - \id \right) \ket{\psi(t)}.
\end{align}

Here we are interested in a continuous description, where the Poissonian statistics of the jumps $L_k^\prime = L_k + l_k$ become a white noise. Indeed if $|l_k|$ is arbitrarily increased, we can see from Eqs.~\eqref{eq:probs} 
the probability of the jumps $P_k(dt)$ may become comparable to $P_0(dt)$. The continuous limit is achieved when the jumps become very probable, but their effect on the system is very small. We then consider a coarse-grained evolution 
such that many jumps are detected in every single time interval $\Delta t$ but the change in the system is still infinitesimal (see Ref.~\cite{Wiseman93}), that is $\Delta t ~\simeq \epsilon^{3/2} \ll 1$ and $|l_k| \simeq \epsilon^{-1}$.
In this case the central limit theorem can be applied, and the probability distribution for the number of jumps $\Delta N_k$ of type $k$ during $\Delta t$ becomes Gaussian:
\begin{equation} \label{eq:njump}
\Delta N_k  = |l_k|^2 \left( 1 + \frac{\langle X_k\rangle}{|l_k|} + \mathcal{O}(\epsilon^{3/2}) \right) + \Delta W_k |l_k| (1 + \mathcal{O}(\epsilon^{1/2})),
\end{equation}
where $\Delta W_k$ is a Wiener increment verifying $\Delta W_k \Delta W_l = \delta_{k, l} \Delta t$. The (unnormalized) state of the system after $\Delta t$ depends on number of jumps $J$ detected during the interval 
and their precise sequence $\{(t_J, k_J), ..., (t_1,k_1)\}$, that is:
\begin{equation}\label{eq:sequ}
 \ket{\tilde \psi(t + \Delta t)} = U_\mathrm{eff}(\Delta t - t_J) M_{k_J} U_\mathrm{eff}(t_J - t_{J-1})~ ... ~M_{k_1} U_\mathrm{eff}(t_1 - t) \ket{\psi(t)},
\end{equation}
where $M_{k_j}$ are the operators in Eq.~\eqref{eq:jumps} ($k_j \neq 0$) and $U_\mathrm{eff}(t-s) \equiv \exp\left[-i(H + \frac{i}{2} \sum_k L_k^{\dagger} L_k + |l_k| X_k + |l_k|^2)(t-s)\right]$ describes the smooth evolution periods where no jumps are detected.
Here it is worth noticing that, on the relevant time-scales ($\epsilon^{3/2}$), the operators in Eq.~\eqref{eq:sequ} commute, so that we can approximate the unnormalized state after the different $\Delta N_k$ jumps of type $k$ as
\begin{equation}\label{eq:ope}
 \ket{\tilde \psi(t + \Delta t)} \simeq U_\mathrm{eff}(\Delta t) \prod_{j= 1}^J M_{k_j} \ket{\psi(t)} \simeq U_\mathrm{eff}(\Delta t) \prod_{k} \left(L_{k} + l_{k}\right)^{\Delta N_k} \ket{\psi(t)}.
\end{equation}
In the following we will assume for simplicity that $\varphi_k = 0$ for all $k$. Expanding Eq.~\eqref{eq:ope} in orders of $\epsilon$ and keeping terms up to $\epsilon^{3/2}$ we obtain:
\begin{equation}
\ket{\tilde \psi(t + \Delta t)} \simeq \left[1 - iH \Delta t - \frac{ \Delta t}{2}\sum_k \left(L_k^\dagger L_k - \langle X_k \rangle L_k \right) + \sum_k\Delta W_k L_k \right] \ket{\psi(t)},
\end{equation}
where we neglected a multiplicative term $l_k^{\Delta N_k} \exp[-|l_k|^2 \Delta t]$ irrelevant for the unnormalized state $\ket{\tilde{\psi}(t + \Delta t)}$. Now taking the limit $l_k \rightarrow \infty$, so that $\epsilon \rightarrow 0$, we can replace $\Delta t$ by $dt$ and $\Delta W_k$ by $dW_k$. 
The stochastic Wiener increments $dW_k(t)$ represent a white noise contribution, such that $d W_k d W_l = \delta_{k, l} d t$ and average over trajectories $\langle dW_k \rangle = 0$. Including normalization, we obtain the final form of the diffusive 
stochastic Schr\"odinger equation:
\begin{equation}\label{eq:stoch}
d\ket{\psi(t)} = \left[-i H dt - \frac{dt}{2} \sum_k \left(L_k^\dagger L_k - L_k \langle X_k \rangle_{\psi(t)} + \frac{1}{4}\langle X_k \rangle_{\psi(t)}^2 \right) + \sum_k dW_k(t) \left( L_k - \frac{\langle X_k \rangle_{\psi(t)}}{2} \right) \right] \ket{\psi(t)},
\end{equation}
which, upon identifying $H_\mathrm{eff} \equiv H - \frac{i}{2}\sum_k L_k^\dagger L_k$, matches the form reported in Eq.~(2) of the main text.
The output currents associated to the measurements can be obtained by removing the constant displacement from the signals in Eq.~\eqref{eq:njump}, and taking the continuous limit:
\begin{equation}
 J_k(t) \equiv \lim_{l_k \rightarrow \infty} \frac{dN_k(t) - l_k^2 dt}{l_k dt} = \langle X_k\rangle_{\psi(t)} + \xi_k(t),
\end{equation}
where $\xi_k(t) \equiv dW_k(t)/dt$, corresponding to a continuous measurement of the quantity $X_k$.

Finally, from Eq.~\eqref{eq:stoch} we can calculate the corresponding stochastic master equation for the conditioned density operator $\varrho(t)\equiv \ket{\psi(t)}\bra{\psi(t)}$. It reads:
\begin{equation} \label{eq:stomas}
 d\varrho(t) = -i[H,\varrho(t)]dt + \sum_k \mathcal{D}[L_k](\varrho) dt + \sum_k \mathcal{H}[L_k](\varrho) dW_k(t),  
\end{equation}
where we introduced the measurement superoperator $\mathcal{H}[L](\varrho)= L\varrho + \varrho L^\dagger - \langle X_k \rangle_{\varrho(t)} \varrho$. We notice that the above stochastic master equation is in the general form reported 
in Refs.~\cite{Jacobs,Diosi} for ideal (efficient) detectors. Taking the average over trajectories, we can easily verify that since $\langle dW_k(t)\rangle = 0$, Eq.~\eqref{eq:stomas} reduces to the standard master equation~(1) 
of the main text.

\end{document}